\documentclass[english,aps,preprint]{revtex4}
\usepackage{graphicx}
\usepackage[T1]{fontenc}
\usepackage[latin1]{inputenc}
\usepackage{amssymb}

\makeatletter

\providecommand{\LyX}{L\kern-.1667em\lower.25em\hbox{Y}\kern-.125emX\@}

\usepackage{babel}
\makeatother
\newcommand{\beq}{\begin{equation}}
\newcommand{\eeq}{\end{equation}}

\newcommand{\etab}{\mbox{\boldmath $\eta $}}
\newcommand{\bPi}{\mbox{\boldmath $\Pi $}}

\newcommand{\barrho}{{\bar{\rho}}}

\def\bA{{\mathbf A}}

\def\bp{{\mathbf p}}

\def\bI{{\mathbf I}}

\def\bq{{\mathbf q}}
\def\br{{\mathbf r}}

\def\bS{{\mathbf S}}

\def\bR{{\mathbf R}}
\def\bz{{\mathbf z}}

\def\bk{{\mathbf k}}

\def\l{{\lambda}}
\def\G{{\Gamma}}

\def\a{\alpha}

\def\half{{1\over2}}

\def\intq{{\int {d^2q\over(2\pi)^2}}}
\def\intk{{\int {d^2k\over(2\pi)^2}}}

\def\hz{{\hat \bz}}
\def\ua{\uparrow}
\def\da{\downarrow}
\def\beqr{\begin{eqnarray}}
\def\eeqr{\end{eqnarray}}
\def\ddag{{d^{\dagger}}}

\def\prl{{Phys. Rev. Lett.}}
\def\prb{{Phys. Rev. {\bf B}}}

\def\rmp{{Rev. Mod. Phys.}}
\begin{document}

\title{Hamiltonian theory of the half-filled Landau level with
disorder: Application to recent NMR data}

\author{Ganpathy Murthy}

\email{murthy@pa.uky.edu}

\affiliation{Department of Physics and Astronomy, University of
Kentucky, Lexington KY 40506-0055}

\author{R. Shankar}

\email{r.shankar@yale.edu}

\affiliation{Department of Physics, Yale University, New Haven, CT
06520}

\begin{abstract}
The Hamiltonian Theory of the fractional quantum Hall effect is an
operator description that subsumes many properties of Composite
Fermions, applies to gapped and gapless cases, and has been found to
provide results in quantitative accord with data on gaps, relaxation
rates and polarizations at temperatures of $300mK$ and above.  The
only free parameter is $\lambda$, which is related to the sample
thickness and appears in the Zhang-Das Sarma potential $v(q) = {2\pi
e^2\over \kappa q} e^{-ql\lambda}$ where $l$ and $\kappa $ are the
magnetic length and dielectric constant. Here we examine the recent
data of Tracy and Eisenstein on the nuclear magnetic resonance
relaxation rate at filling factor $\nu=\half$ deduced from resistivity
measurements at temperatures as low as $45mK$. We find that their
results can be satisfactorily described by this theory, if in addition
to a $v(q)$ with $\lambda \simeq 2$, a constant disorder width
$\Gamma\simeq 100 mK$ is incorporated.
\end{abstract}
\maketitle

\section{introduction}

There are several theoretical approaches to the  fractional
quantum Hall effect (FQHE)\cite{fqhe-ex,perspectives}. For
fractions of the form  $\nu=\frac{1}{2p+1}$ one can write down
inspired trial wave functions following  Laughlin\cite{laugh} to
describe the incompressible liquid ground state and its gapped
excitations. These  excitations can be further studied using an
extension of the Bijl-Feynman approach of Girvin, MacDonald and
Platzman \cite{GMP}.

For a more general class of fractions with filling $\nu=p/(2ps+1)$, one can
follow Jain\cite{jain-cf,jain-cf-review} and work with Composite
Fermions (CF) which are obtained from electrons by attaching $2s$ flux
quanta \cite{cs-trans,gcs}. On average, these attached flux quanta
neutralize enough of the external flux so that the total is just right
for the CF's to fill exactly $p$ Landau levels. When the wave
functions of these unique states are gauge transformed back to
electronic language and projected to the Lowest Landau Level (LLL) of
electrons, they become excellent trial wave functions. The
quasiparticle excitations can likewise be described by starting with
particle-hole excitations of these $p$-filled CF-LL's (Composite
Fermion Landau levels).

For the Laughlin fractions Zhang, Hansson, and Kivelson \cite{zhk}
provided a microscopic Chern-Simons (CS) field theory directly linked
to the original Hamiltonian.  They converted electrons to bosons by
adding $2p+1$ flux quanta so that on average the bosons saw no flux.
Lopez and Fradkin \cite{lopez} extended the CS field theory to the
Jain fractions $p/(2ps+1)$. In both the bosonic and fermionic cases
fluctuations on top of the mean field description were described a CS
gauge field.  The path integral approach allowed one to more readily
compute time and space dependent response functions and go to nonzero
temperature $T$.

The CS theory applied  to $\nu=\frac{1}{2}$ leads to CF's that
see, on average,
zero effective field.  Kalmeyer and Zhang\cite{kalmeyer}, and
Halperin, Lee, and Read (HLR)\cite{hlr} studied the Fermi sea of
CF's that coupled to CS gauge field.  The latter authors performed
an exhaustive and definitive study making predictions in agreement
with many experiments.

However all the CS theories have one weakness: They do not yield a
smooth limit as one projects to the LLL by sending $m \to 0$. This
sends the electronic cyclotron frequency $\omega_c = {eB\over
mc}\to\infty$, thereby forcing all the electrons to be in the LLL (for
$\nu\le 1$). We know that a smooth limit must exist, and that in this
limit the kinetic energy should become an ignorable constant and the
entire Hamiltonian should be just the Coulomb interaction $v(q)$
between electrons. The quasiparticle mass, its band structure, its
residual interactions, response functions, and so on should be
determined solely by $v(q)$. In the Chern-Simons theories sending $m$
to zero causes unavoidable singularities. These theories also lead at
the mean-field level to a CF of charge $e$ and not the correct quasiparticle charge of
\beq e^*= {e \over
2ps+1}= e(1-c^2)
\eeq
where 
\beq c^2 = {2ps\over 2ps+1}
\eeq
is a constant that will appear repeatedly in this paper.

 The electric charge does not come out right because the CS procedure
attaches the flux but not the screening charge associated with the
nucleation of a $2s$-fold vortex so evident in the wave function
description\cite{laugh,jain-cf}. (In other words, the flux captures
only the phase of the vortices but not their modulus, with its
$2s$-fold zero.)  In particular, at $\nu = 1/2,\ c^2=1$, it does not
yield a neutral fermion or describe its dipole structure \cite{read2}.

The present authors addressed these problems by developing a
Hamiltonian approach\cite{us1,us2,aftermath,rmp-us} in which the LLL
limit can be taken naturally, and many properties of the CF, such as
its reduced charge $e^*=1/2ps+1$, and its dipolar nature are encoded
unambiguously in the operator structure. Relegating the details of
this theory to the next section, we merely point out that it not only
reproduces results found by Monte Carlo calculation based on trial
wave functions \cite{park-jain}, (for potentials that are not too
singular at short distances), it also furnishes results in
quantitative accord with data on gaps, relaxation rates and
polarizations at temperatures of $300mK$ and above for gapped and
gapless fractions \cite{shankar-NMR,pol-me}.

  The sense in which the Hamiltonian theory accounts for the data
needs some elaboration\cite{shankar-NMR}. Consider the $\nu =1/2$ data
of Dementyev {\em et al}\cite{dementyev}, and Khandelwal
\cite{khandelwal}, who measured the polarization and the NMR
relaxation rate for relatively high temperatures (down to
$300mK$). One can easily find a simple CF theory with an effective
mass, and if needed, a Stoner interaction $J$, to fit the data in a
limited range. There are two shortcomings in this approach.  First,
the origin of the CF mass and interaction are viewed as unrelated when
in fact they both originate from the electrostatic interaction between
electrons.  Next, as emphasized by Dementyev {\em et
al}\cite{dementyev} and Khandelwal\cite{khandelwal}, if one considers
all of their data globally, for both polarization and relaxation, the
values of $(m,J)$ needed for one observable are totally incompatible
with those mandated by the other.  This is because no single pair
$(m,J)$ can describe the underlying physics.

On the other hand, in the Hamiltonian theory, the only free parameter
is $\lambda$, a parameter related to the sample thickness in the
Zhang-Das Sarma potential\cite{zds}
\beq v(q) = {2\pi e^2\over \kappa
q} e^{-ql\lambda} 
\eeq
where $l=\sqrt{{\hbar c\over eB}}$ is the magnetic length, and
$\kappa$ is the dielectric constant. The parameter $\lambda$ is
extracted from one data point, after which the theory does quite well
at predicting the rest of the data\cite{shankar-NMR}.  As we review
the theory, it will become apparent how the CF band structure and
interactions (whose nonstandard functional forms are fully determined
by $v(q)$) will arise in a correlated way, and vary with the chemical
potential, temperature etc., to produce varying effective mass and
interaction parameters.

The current work was triggered by recent experiments of Tracy and
Eisenstein (TE)\cite{TE}.  They measured the NMR relaxation rate
$\frac{1}{T_{1}}$ at much lower temperatures than previously (going as
low as 45 mK) and for a wide range of $B$ fields while keeping the
filling factor at $\nu=\frac{1}{2}$. The relaxation rate was measured
by disturbing the system magnetically, and measuring the relaxation of
the resistance to its equilibrium value. The precise causal
relationship between the non-equilibrium value of polarization and the
non-equilibrium value of the resistance is obviously not needed to
extract their common relaxation rate.

We do not have much to say about the region of very high  $B$
where the polarization has saturated and the spin up and down
bands have separated. Here we simply expect the following
asymptotic behavior:
\beq T_{1}(B,T)\approx\exp{\frac{E_{Z}-E_{Z}^{*}}{T}}
\label{exprise}
\eeq
where $E_{Z}$ is the Zeeman coupling, and $E_{Z}^{*}$ is the
critical Zeeman coupling at which the system becomes fully
polarized.

Our focus is on the region before saturation where the two spin bands
overlap.  What does a naive model of noninteracting CF's predict here?
One can easily argue that:
\beq {1 \over T_{1}T}\simeq |u(0)|^4 m^{*2} \approx B^{5/3}
\label{13/6}\eeq
where $m^*$ is the effective mass (and controls the density of states)
and where $u(0)$ measures the height of the wave-function in the
transverse direction.  Two factors contribute to the final power of
$B$: (i) The density of states per unit volume for each spin species
scales as $m^*\simeq B^{1/2}$ (which in turn follows from setting
$k_{F}^{2}/m^*\simeq e^2/l$) (ii) The factor $|u(0)|^4$, scales as
$B^{2/3}$ because the thickness $t$ of the 2DEG scales as
$n^{-1/3}\simeq B^{-1/3}$\cite{ando-fowler-stern} and $|u(0)|^2$ integrates to unity in the
transverse coordinate.
Thus one expects that at fixed $T$, 
\beq
 T_1 \simeq B^{-5/3}.
\eeq
However the TE data   show a nearly flat dependence on $B$
till saturation begins to set in, in contradiction with naive CF theory.

In the Hamiltonian theory the $B$ dependence of $T_1$ is much more
complicated.  First the density of states (and hence $m^*$) varies
with the chemical potential since the bands are not quadratic.  Next
the self-consistent band structure depends on another energy scale,
the Zeeman energy, and on the thickness parameter $\lambda$, which
scales with $B$ as $t/l_{0}\simeq B^{1/6}$. All of this precludes any
naive scaling.  One simply has to turn the crank to see how things
depend on $B$. We did that and here are the main results.

\begin{figure}[t]
\includegraphics[height=5in,width=5in]{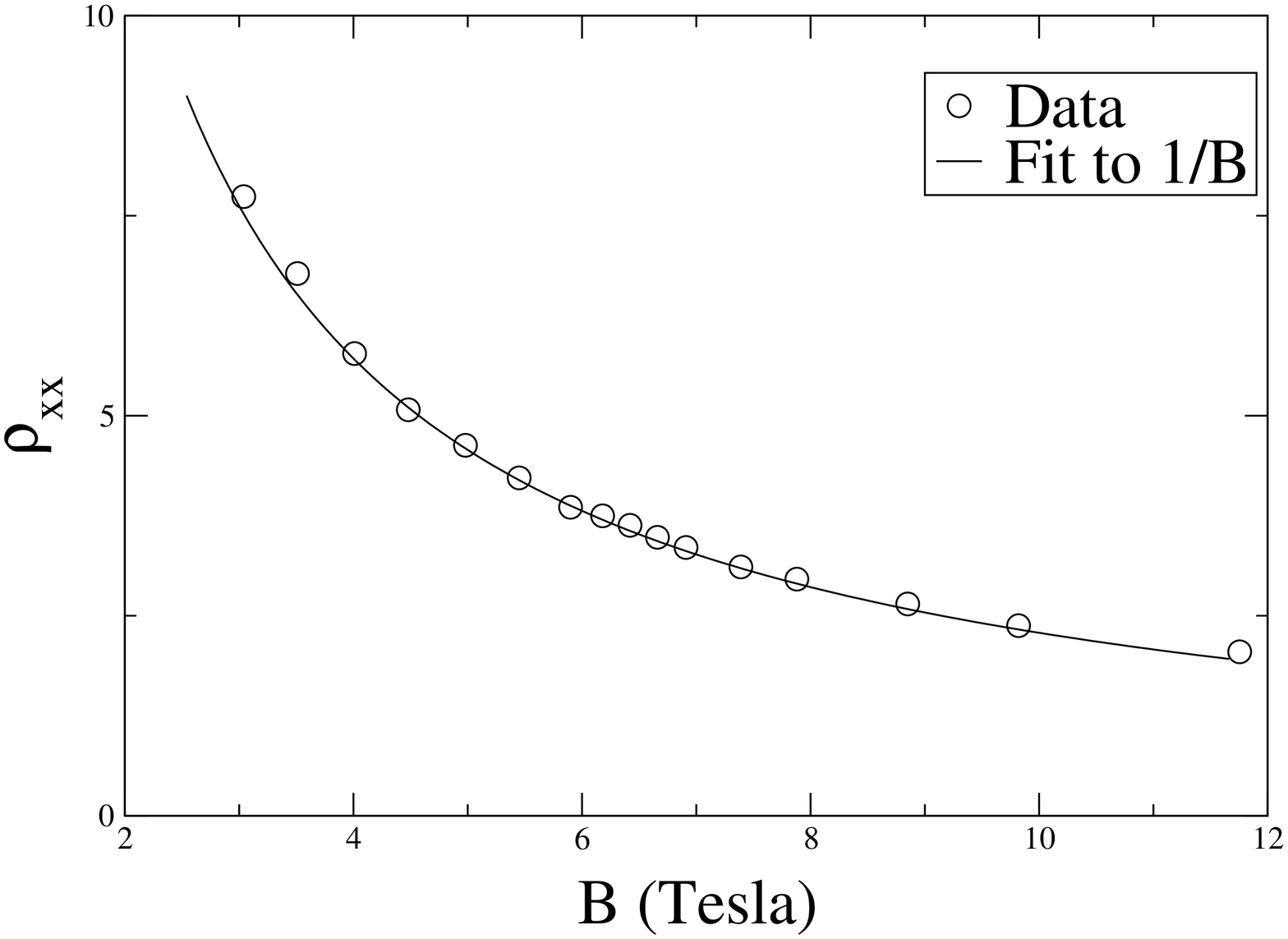}
\caption{The data from Tracy and Eisenstein, Ref. \cite{TE},
fitted to a $1/B$ curve. The good fit implies that the transport
time is independent of $B$ or $n$.} \label{fig1}
\end{figure}

We found that while a suitable $\lambda \simeq 2$ (at $4\
\mbox{Tesla}$) could produce acceptable graphs, the data clearly
called for a nonzero disorder-induced width $\Gamma$ of order $100mK$
for all momentum states.  The existence of such a $\Gamma$ and its order of
magnitude both are implicit in TE data for the equilibrium
resistivity\cite{TE}, reproduced in Fig. (\ref{fig1}).  Assuming a
Drude form for the conductivity
\beq \sigma=\frac{ne^{2}\tau}{m^{*}} \label{drude}\eeq
with $e$ being the {\it electronic charge}, and $m^*$ being the
effective mass of the CF's, one finds that the disorder width
$\Gamma=\frac{\hbar}{2\tau}$ is a fraction of a Kelvin. Assuming that
$\Gamma$ is roughly independent of $n$ and therefore $B$, fits the
data quite well, as seen in Fig. (\ref{fig1}).  In past comparisons to
data at $300mK$ and higher\cite{shankar-NMR}, disorder played a
relatively minor role presumably because the temperature was much
larger than the $\G$ of those samples.

In this paper our goal is limited to understanding the effect of
disorder on the NMR relaxation rate, and we merely take Eq.
(\ref{drude}) as a crude estimate of disorder broadening and do not
attempt to provide a complete theory of longitudinal conductance at
$\nu = 1/2$, a subject that is yet to be fully developed. (See
\cite{hlr} for a treatment of the charged Chern-Simons fermion coupled
to a gauge field.)

\begin{figure}[t]
\includegraphics[height=3in,width=3in]{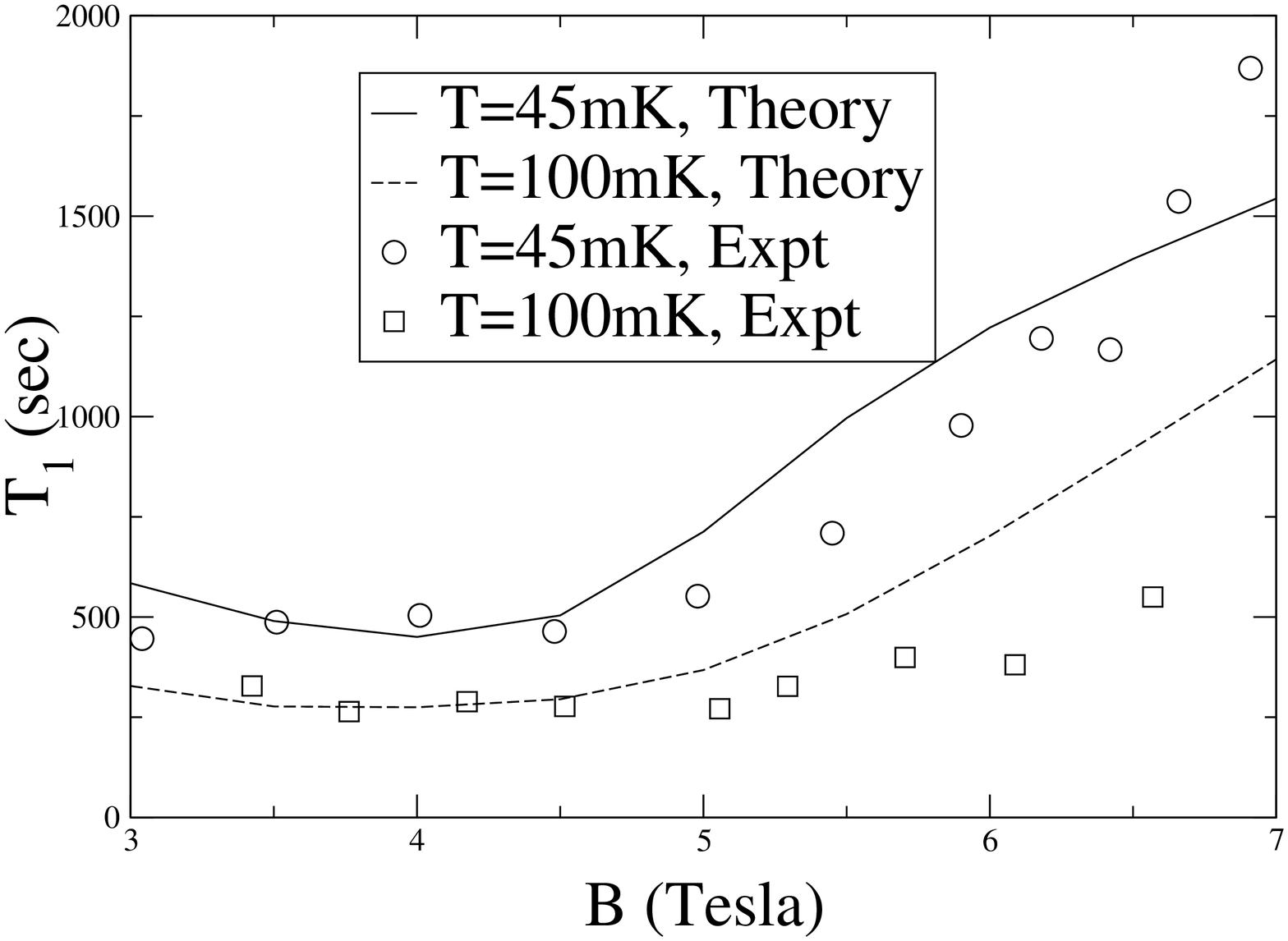}
\caption{Comparison of theory and experiment  for  $T_1$ vs $B$ at
$45mK$ (solid line, circles) ) and $100mK$ (dotted line, squares).
The optimal values  $\lambda = 2$ (at $4T$)  and $\Gamma=100mK$
were employed. In the comparison, an ``offset'' of
$5\times10^{-4}s^{-1}$ has been added to the nuclear relaxation
rate to account for the diffusion of nuclear spins out  2DEG into
the bulk. }\label{fig2}
\end{figure}

Figure \ref{fig2} shows a comparison of our theory to the $T_1$ data
at 45mK and 100mK. We find that the numerical values $\lambda=2.00$
and $\Gamma=100mK$ produce the best overall agreement. A
$B$-independent parallel relaxation channel, with rate
$5\times10^{-4}s^{-1}$, (representing nuclear spin diffusion out of
the 2DEG, and estimated from the TE data), has been added to our
theory, and our prediction has been normalized to coincide with the
data at $B=4\ Tesla$. Finally $\l (B)$ was scaled from its value at $4\
Tesla$.

\begin{figure}[b]
\includegraphics[height=3in,width=3in]{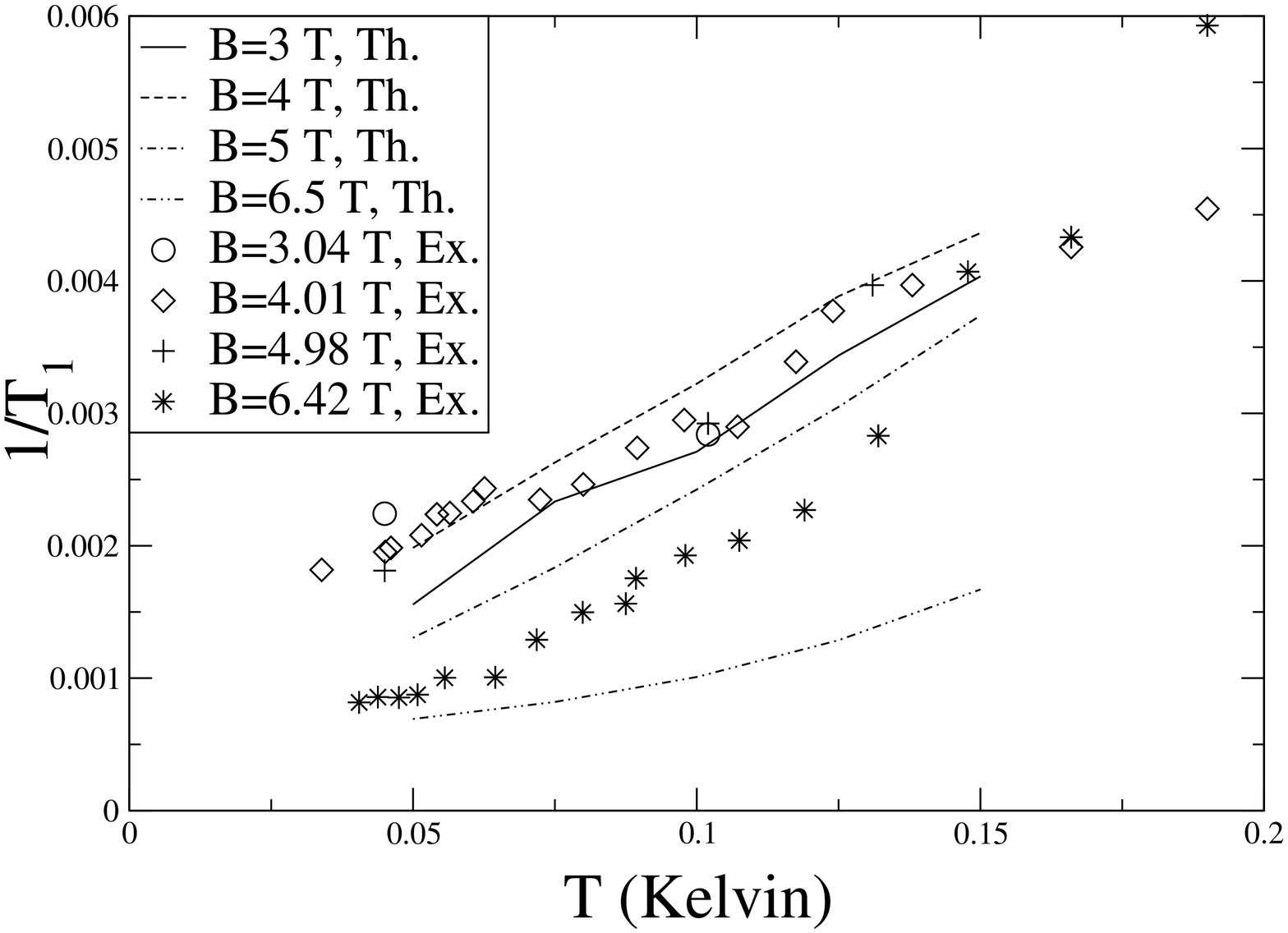}
\caption{Comparison of theory and experiment for $1/T_1$ vs $T$ In the
 comparison, an ``offset'' of $5\times10^{-4}s^{-1}$ has been added to
 the nuclear relaxation rate to account for the diffusion of nuclear
 spins out 2DEG into the bulk semiconductor. The parameters used were
 $\lambda = 2, \Gamma= 100mK$.  }\label{fig3}
\end{figure}
Disorder seems to mimic the effects of nonzero temperature in many
ways. We see a clear flattening of the results for $T_{1}$ below
$E_{Z}^{*}$. Above $E_{Z}^{*}$, since the bottom of the minority
spin band is disorder-broadened, there is always some density of
states of both spin species near the chemical potential, and
therefore $T_{1}$ rises much slower than the activated form of Eq.
(\ref{exprise}).

Figure \ref{fig3} shows the comparison of theory versus experiment
for $1/T_1$ versus $T$ for several $B$ values. As at low $T$, the
theory shows more $B$-dependence than is measured, and does not do
too well past saturation.

The plan of this paper is as follows.  In Section II we give a brief
review of the Hamiltonian Theory. In Section III we introduce our
constant width disorder model, set up the HF equations, and present a
formula for the NMR relaxation rate and polarization.  Section IV is
devoted to a comparison of theory to experiment and the role of our
two free parameters, and of the temperature, on the NMR relaxation
rate. In Section V we address the effects of Landau-level mixing
qualitatively, and in Section VI we conclude by presenting some of the
limitations of our work, how they may be overcome, and some open
questions.

\section{Review of the Extended Hamiltonian Theory}

We shall furnish only  a brief summary of the Hamiltonian theory
since complete details may be found  in our review
article\cite{rmp-us}. Our starting point will be the full
electronic Hamiltonian

\beq H=\sum\limits_{i} {\bPi_{ei}^2\over 2m}+\int
{d^2q\over(2\pi)^2}v(q)\rho_e(\bq)\rho_e(-\bq) \label{hfull}\eeq

where $m$ is the band mass of the electrons, $\bPi_e=\bp_e+e\bA$
is the velocity operator of the electrons of charge $-e$, $v(q)$
is the Coulomb interaction between the electrons, and
$\rho_e(\bq)=\sum_i\exp{-i\bq\cdot\br_{ei}}$ is the electronic
density operator.

To facilitate the projection to the LLL, we diagonalize the
kinetic energy by decomposing the electron coordinates and momenta
into cyclotron ($\etab_e$) and guiding center (${\bf R}_e$)
variables
\beqr
\etab_e=&l^2\hz\times\bPi_e\\
\bR_e=&\br_e- l^2\hz\times\bPi_e \eeqr

where $l= \sqrt{\hbar c/eB}$ is the magnetic length. These
coordinates obey the commutation relations
\beq \left[ \etab_{ex} \ , \etab_{ey} \right] = il^2 \ \ \ \left[
\bR_{ex} \ , \bR_{ey} \right]=-il^2 \ \ \ \ \left[ \etab_e \ ,
\bR_e \right]=0 
\eeq
The kinetic energy depends only on the
cyclotron coordinate, with the LLL corresponding to the harmonic
oscillator ground state for this variable.  Expressing the
electron coordinate as $\br_e=\bR_e+\etab_e$ we can now take the
LLL limit of the Hamiltonian to get 
\beq {\bar H}_e=\int
{d^2q\over(2\pi)^2}v(q)\barrho_e(\bq)\barrho_e(-\bq)
\label{barHe}\eeq
where $\barrho_e$ is the electron density
operator projected to the LLL. 
\beq \barrho_e(\bq )=e^{-{1 \over
4}(ql)^{2}}\sum\limits_{i}e^{-i\bq\cdot\bR_{ei}}. \eeq 
While we
can easily project the Hamiltonian to the LLL, working with it is
another thing.   Here, briefly, are the problems we face and their
resolution. The original electronic problem assigns to each
particle two coordinates $(x_e,y_e)=\br_e $ and two conjugate
momenta $(p_{ex},p_{ey})=\bp_e$. In the LLL, kinetic energy is
quenched but $x_e$ and $y_e$ become conjugate variables so that
$H=v(x_e,y_e)$ poses  a nontrivial quantum problem.

Equivalently, each electron has a cyclotron coordinate $\etab_e$
and guiding center coordinate $\bR_e$. In the LLL the Hamiltonian
is $v(\bR_e)$ with $R_{ex}$ and $R_{ey}$ being conjugate
variables.  Quantizing such a problem is tricky and involves
working with analytic wave functions \cite{girvin-jach}.

Our trick was to begin with   new particles, the CF's, equal in
number to the electrons, each  with coordinate $\br$ and velocity
$\bPi =\bp +e^*\bA$, where $e^*=e/(2ps+1)$ is the CF charge.  Thus
$\bPi$ describes particles which will fill exactly $p$ LL's. In
what follows we will also work with $\etab$ and $\bR$, the
corresponding cyclotron and guiding center coordinates of the CF.
Note that the CF variables do not carry subscripts.

In this space we
 form the following entities:
\beqr
\bR_e=&\bR+\etab c=\br-{l^2\over1+c}\hz\times\bPi\\
\bR_v=&\bR+\etab/c=\br+{l^2\over c(1+c)}\hz\times\bPi \eeqr
The first vector $\bR_e$ is identified with the electronic guiding center
coordinate since it obeys 
\beq \left[ R_{ex}\ ,
R_{ey} \right] =  -il^2. 
\eeq
Thus the LLL projected electronic Hamiltonian we set out to solve
is just $H=v(\bR_e )$. Since a quantum problem is characterized by
the commutation rules, this is a faithful transcription of our
original mission. However, there is no assurance the degeneracy of
our levels will be the same now.

Indeed we can see there is going to be a huge degeneracy because
of the other coordinate $\bR_v$  It describes a particle we call
the pseudo-vortex. It commutes with $\bR_e$: 
\beq \left[ {\bf
R}_e\ , {\bf R}_{v} \right]= 0.
\eeq
  and obeys
\begin{eqnarray}
\left[ R_{vx}\ , R_{vy} \right] &=&  il^2/c^2,
\end{eqnarray}
Thus $\bR_v$ is a cyclic variable whose dynamics is unrelated to
the original problem. Let us get acquainted with it anyway, since
all this will change.

First of all, it  follows from its commutation rules that $\bR_v$ 
describes a particle of charge $-c^2$ in electronic units. If this
object paired with an electron, it would yield an entity with
total charge $e(1-c^2)= e/(2ps+1)=e^*$. This is exactly how the
vortices in the trial wave functions screen the electron to give
rise to the CF\cite{jain-cf}. However the adjective "pseudo" is appended for two
reasons: (i) The vortices in the wave functions are not creatures
with their own coordinates independent of electrons. (ii) The
pseudo-vortex so far has nothing to do with the original problem,
but is rather something we introduced when we enlarged the Hilbert
from the LLL projected case described by just $\bR_e$ to a regular
fermionic space, i.e., the CF space.

 Since $H=v(\bR_e )$ does
not depend on $\bR_v$, the dynamics of $\bR_v$ is unspecified.
They are like gauge variables and their dynamics is determined by
gauge fixing. We made the reasonable choice  that the density
$\rho_v$  formed out of $\bR_v$ annihilated all physical states.

This gives us two options.

The first is to start with a self-consistent Hartree-Fock (HF)
solution to the Hamiltonian of Eq. (\ref{barHe}) written in CF
coordinates, and to compute response functions in a ``conserving''
approximation\cite{kada-baym}, such as time-dependent Hartree-Fock
(TDHF). In this approximation  the constraints are satisfied are
satisfied ``weakly'', in  correlation
functions\cite{read3,conserving-me}.

The second approximation, which we have employed in the
past\cite{us1,us2,pol-me,shankar-NMR}, and use in this manuscript, is
more unconventional. We argue that when acting on exact physical
states (which are annihilated by $\barrho_v$) there should be no
difference between the projected electron density operator
$\barrho_e$, and the following {\it preferred density operator}
\beq \barrho_p(\bq)=\barrho_e(\bq)-c^2\barrho_v(\bq). \eeq
The new operator $\barrho_p$ has the advantage of exhibiting many of
the nonperturbative properties of the CF without any computation, at
the tree level. First, it describes a particle of charge
$e^*=e(1-c^2)$ as can be seen by looking at the zeroth order term in
an expansion $\rho_p (q) $ in powers of $q$. The order $q$ term has
the correct coefficient to impose another crucial property dictated
by Kohn's theorem\cite{kohn}, that any intra-LLL matrix element of the
density operator should vanish faster than linear in $q$ as
$q\to0$. This property emerges from the conserving calculation only
after some effort\cite{read3,conserving-me}. It is very striking that
one and the same admixture of $\rho_v$, with coefficient $c^2$, serves
two purposes at once.

As mentioned this situation is unusual. Normally we have a fixed
Hamiltonian and go in search of a trial HF state from a family, here
we have a fixed HF state (with $p$ filled LL of CF's) and go searching
among a set of Hamiltonians equivalent in the constrained space. If
the constraint is solved exactly, there is nothing to choose between
the original $\barrho_e$, the preferred $\barrho_p$, or indeed, any
arbitrary combination of $\barrho_e(\bq)$ and $\barrho_v(\bq)$. On the
other hand in HF, the constraint is not satisfied and the preferred
combination emerges as the best to use since it encodes certain
important nonperturbative properties of the CF at leading order.

We will henceforth use the preferred form of the density in all
our calculations, to which we now turn.

\section{NMR Relaxation Rate for $\nu=\half$}

The assumption that all the nuclei are in thermal equilibrium with
each other on a fast time-scale leads to the standard Korringa Law
for nuclei in contact with degenerate Fermi gases\cite{slichter}.
The formalism for the clean $\nu=\half$ system has been previously
worked out by one of us\cite{shankar-NMR}. We will proceed
directly to the case with disorder. Our model Hamiltonian is the
following:
\beq H=\intq v(q) \barrho_p(\bq)\barrho_p(-\bq)-E_ZS_z+H_{dis}
\label{Hpref}\eeq
where $E_Z$ is the Zeeman energy, $H_{dis}$ is the coupling to
disorder, and we explicitly present our preferred density operator
for $\nu=\half$
\beq
\barrho_p(\bq)=e^{-(ql)^2/4}\sum\limits_{\bk}-2i\sin({\bq\times\bk
l^2\over2}) \ddag_s(\bk)d_s(\bk) \eeq
Here $\ddag_s(\bk)$ creates a CF in a state with momentum $\bk$
and spin-projection $s$.  For our simple model, we do not need to
specify the detailed form of the coupling to disorder. We assume
that the net effect of $H_{dis}$ in a disorder-averaged treatment
is to provide a momentum- and spin-independent width $\Gamma$ to
every single-particle state. We incorporate this into our HF
energy calculation as follows. Let the disorder-averaged energy of
a state labelled by $\bk,s$ be $\epsilon_s(\bk)$, and its
occupation be $n_s(\bk)$. Then decomposing the Hamiltonian of Eq.
(\ref{Hpref}) in the HF aproximation, we find
\beqr \epsilon_s(\bk)=&-E_Z{s\over2}+\intq
v(q)e^{-(ql)^2/2}[1-\cos(\bq\times\bk
l^2)][1-n_s(\bk+\bq)-n_s(\bk-\bq)] \\
&n_s(\bk)=\int\limits_{-\infty}^{\infty} d\epsilon'
{\Gamma\over\pi(\epsilon'^2+\Gamma^2)}
{1\over1+e^{\beta(\epsilon_s(\bk)-\epsilon'-\mu)}}
\label{HFeqs}\eeqr
where $\beta=1/k_BT$ is the inverse temperature. The form of the
first equation is the same as in the clean system, while the
disorder-broadening $\Gamma$ makes itself felt in the occupations.
It says that since a particle in a state of momentum $\bk$
(defined after averaging) can be in a state of energy different
from $\varepsilon(\bk )$ by an amount of order $\Gamma$, a band of
energy states of width $\Gamma$ will contribute to $n(\bk )$ by
convolution.

The spectral function can be related to the disorder average of
the exact disorder eigenstates in the HF basis. Let the exact
disorder eigenstates be labelled by $\a$. We can expand the
operators which annihiliate a CF in a momentum state in terms of
$\a$:
\beq d_s(\bk)=\sum\limits_{\a} \phi_{\a}(\bk) d_{s\a} \eeq
Now the retarded single-CF Green's function can be defined as
\beq
G_{s}^{R}(\bk,\bk',t)=-i\Theta(t)\langle\Omega|\{d_s(\bk,t),\ddag_s(\bk',0)|\Omega\rangle
\eeq
Fourier transforming and expanding in the exact disorder states we
find
\beq G_{s}^{R}(\bk,\bk',\omega)=\sum\limits_{\a}
{\phi_\a(\bk)\phi^*_\a(\bk')\over\omega-E_\a+i0^+} \eeq
Now taking the disorder average and comparing to the spectral
function, we infer that
\beq
\overline{\phi_\a(\bk)\phi^*_\a(\bk')}=\delta_{\bk\bk'}{\Gamma/\pi\over\Gamma^2+(E_\a-\epsilon_s(\bk))^2}
\eeq
 When Eqs. (\ref{HFeqs}) are iterated to
self-consistency they provide the HF state of the CF-Fermi sea in
the presence of disorder. Of course, one needs to maintain the
half-filling condition at every HF iteration: \beq \intk
[n_\ua(\bk)+n_\da(\bk)]={1\over4\pi l^2} \eeq

Now one needs to address the spin correlations, which are
important for the NMR relaxation rate\cite{shankar-NMR}. The
hyperfine interaction between the electrons and the nuclei is
given by \beq {8\pi\over3}\gamma_e\gamma_n\hbar2
|u(0)|^2\bI\cdot\bS_e(0) \eeq where $\gamma_{e,n}$ are the
magnetic moments of the electron and the nucleus, $u(0)$ is the
electronic wavefunction at the nucleus, $\bI$ is the nuclear spin,
and $\bS_e$ is the electronic spin density operator at the
position of the nucleus. We define this operator as follows:
\beq S_{e}^a(\br=0)={1\over L^2}\sum\limits_{\bk,\bq}
e^{-i\bq\times\bk l^2/2-(ql)^2/4}
\ddag_s(\bk-\bq){\sigma_{ss'}^a\over2}d_{s'}(\bk) \eeq
Note that this operator does not commute with the projected
density operator, but has LLL spin-charge commutation
relations\cite{bilayer-original-th}. Now a Fermi golden rule
calculation produces the following expression for the NMR
relaxation rate\cite{shankar-NMR}:
\beq {1\over T_1}=\pi\big({8\pi\gamma_e\gamma_n\over3}\big)^2
|u(0)|^4\int\limits_{-\infty}^{\infty} dt
\langle\Omega|S_{e}^x(\br=0,0)S_{e}^x(\br=0,t)+S_{e}^y(\br=0,0)S_{e}^y(\br=0,t)|\Omega\rangle
\eeq
In our evaluation of the spin-correlation function, we will ignore
vertex corrections, both due to the gauge field (arising from the
imposition of the constraints\cite{read3,conserving-me}) and from
disorder averaging. We have found from our previous work that the
vertex corrections due to the gauge field manifest themselves at
extremely low energies, where they give rise to a collective
overdamped mode, but are unimportant at not too low temperatures.
The vertex corrections due to disorder are important primarily at
small $\bq$ and at energies much smaller than the disorder width
$\Gamma$. However, our correlator involves a sum over all $\bq$,
and the experiments are carried out at a minimum temperature of
$45mK$, which is of the same order as $\Gamma$. Thus, we expect
the effect of disorder vertex corrections to be negligible as
well.

The rest of the calculation is straightforward. Calling the constant
$\pi\big({8\pi\gamma_e\gamma_n\over3}\big)^2 = D$ and performing the
disorder average, we have
\beqr \overline{{1\over T_1}}=&D{|u(0)|^4\over
L4}\sum\limits_{\bk_i\bk_i'}e^{i{l^2\over2}(\bk_1\times\bk_1'+\bk_2\times\bk_2')-{l^2\over4}((\bk_1-\bk_1')^2+(\bk_2-\bk_2')^2)}\times\nonumber\\
&\int\limits_{-\infty}^{\infty}dt
[\overline{\langle\Omega|\ddag_{\ua}(\bk_1,t)d_{\ua}(\bk_2',0)|\Omega\rangle}
\overline{\langle\Omega|d_{\da}(\bk_1',t)\ddag_{\da}(\bk_2,0)|\Omega\rangle}\nonumber\\
+&\overline{\langle\Omega|\ddag_{\da}(\bk_1,t)d_{\da}(\bk_2',0)|\Omega\rangle}\overline{\langle\Omega|d_{\ua}(\bk_1',t)\ddag_{\ua}(\bk_2,0)|\Omega\rangle}]
\eeqr
Expressing the single-particle Green's functions in terms of the
spectral function, and integrating over $t$, we obtain
\beqr \overline{{1\over T_1}}=&D{|u(0)|^4\over
L4}\sum\limits_{\bk_1\bk_2}\int\limits_{-\infty}^{\infty} dE
n_F(E)(1-n_F(E))\bigg({\Gamma\over\pi}\bigg)^2\times\\
&\bigg({1\over(\Gamma^2+(E-\epsilon_{\ua}(\bk_1))^2)(\Gamma^2+(E-\epsilon_{\da}(\bk_2))^2}+\ua\to\da\bigg)
\eeqr
Now we convert the momentum sums into energy integrals by introducing
the densities of states per unit volume $\rho_{\ua\da}(\epsilon)$ of
the $\ua$ and $\da$ spins to get the final expression
\beqr \overline{{1\over T_1}}=&D|u(0)|^4\int dE
dE_{\ua}dE_{\da}e^{-{l^2\over2}(k_{\ua}^2+k_{\da}^2)}I_0(k_{\ua}k_{\da}l^2/2)\rho_{\ua}(E_{\ua})\rho_{\da}(E_{\da})n_F(E)(1-n_F(E))\times\nonumber\\
&{(\Gamma/\pi)^2\over
(\Gamma^2+(E-E_{\ua})^2)(\Gamma^2+(E-E_{\da})^2)}
\label{finalrate}\eeqr
Here $k_{s}=k_s(E_s)$ are to be understood as the result of
inverting the energy-momentum dispersion relation, and $I_0$ is
the modified Bessel function arising from the angular average of
$\exp(\bk_{\ua}\cdot\bk_{\da}l^2/2)$.  It is important to note
that while the $n_F(E)(1-n_F(E))$ factor gives the dominant $T$
dependence at low $T$, there are hidden $T$ dependencies in the
energy-momentum dispersion relations obtained from the
self-consistent HF solution.

\section{Results and Discussion}

Let us begin by asking how we extract the values for $\l$ and
$\G$. Note that $1/T_1$ has the form 
\beq
\bar{{1\over T_1}} = |u(0)|^4 f(T,\l (B), \G )
\eeq

\begin{figure}[t]
\includegraphics[height=3.5in,width=3.5in]{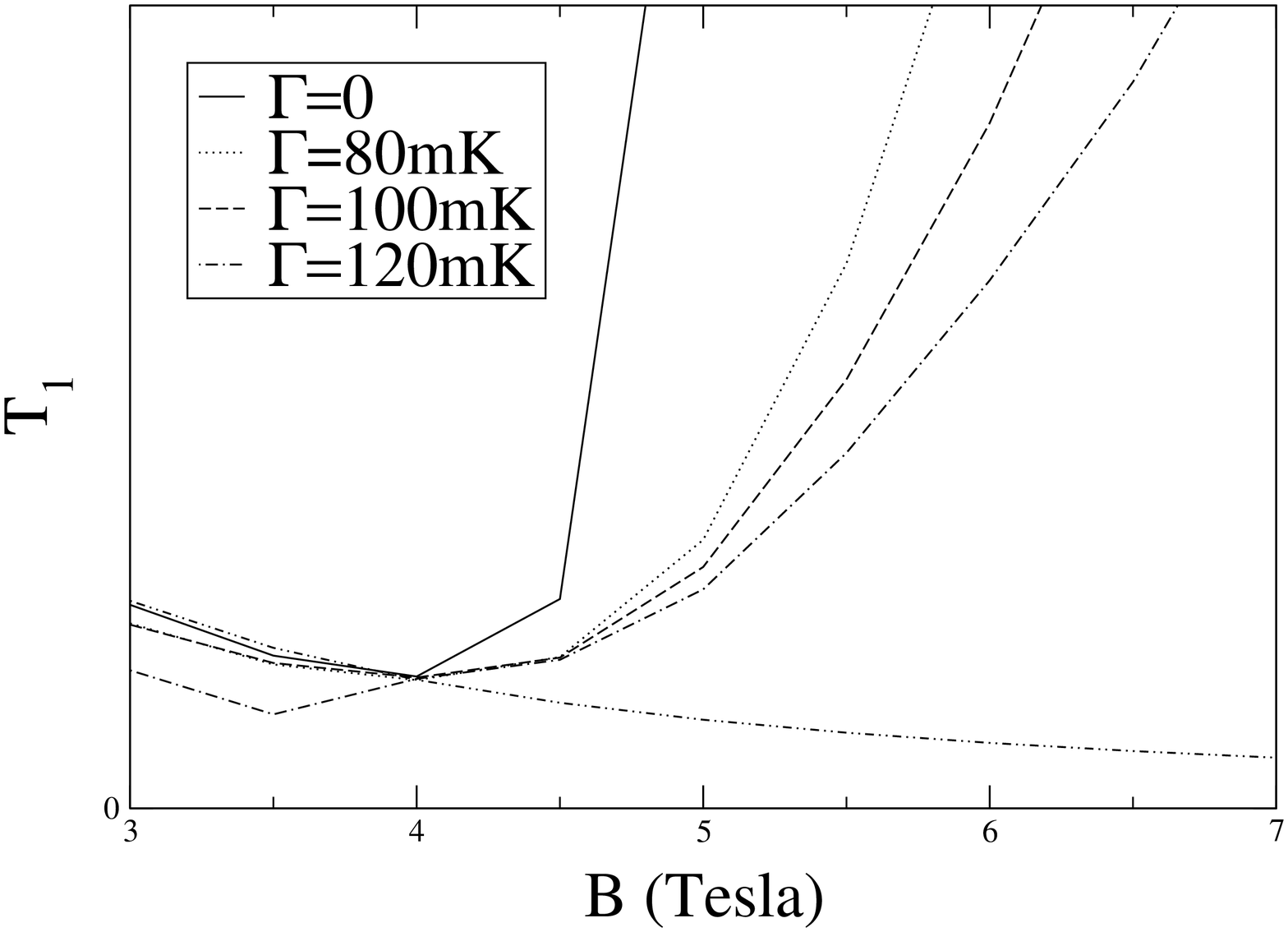}
\caption{A comparison of the $T_1$ vs $B$ curves for
$\lambda=2.00$ for different amounts of disorder. The dashed line
represents the naive CF prediction of $B^{-5/3}$ below full
polarization. $T_1$ is in arbitrary units and the curves have been
normalized to be equal at 4 Tesla for ease of comparison. }
\label{fig4}
\end{figure}

 Let us look at the $T$ dependence of $\bar{{1\over T_1}}$ at
 fixed $B$, say at $4$ Tesla. A rough estimate of $\l$ comes from
 \cite{shankar-NMR} where the following formula is is derived
 \beq
 P= .13 \sqrt{B(T)} \l^{7/4}\ \ \ \  \nu = {1 \over 2}\ \ \ P<1
 \eeq
 relating the polarization to $B(T)$ in Tesla prior to  saturation.
 From TE data which tells us   $P=1$ occurs first for $B(T)\simeq 6.25$ we deduce $\l
 =1.9$ as a good start which must be further improved by looking at the data. The width $\G = 100mK$ is found by considering $T_1$ versus $B$
 data for $T=45\ mK, T=100mK$. None of these numbers is unique in
 that the fit is not perfect and changing the numbers could
 improve things in one region and worsen it in another.
 With these values of $\G$ and $\l$ we pick the overall scale,
 including $|u(0)|^4$, to agree with the data at $4$-Tesla, 45 mK, with the offset of $ 5 \cdot 10^{-4}\ s^{-1}$
 from spin diffusion into the bulk
 included. We are now set to make predictions at any other $B$
 since $\G$ is
 assumed to be constant and the scaling of $\l$ and $|u(0)|^4$ are known:

\beq \lambda(B)=\lambda(4 Tesla) \bigg({B\over 4\ Tesla}\bigg)^{1/6} \eeq
Fig. \ref{fig4} shows the effect of $\Gamma$ when $\lambda$ is
held fixed at $2.00$ and the temperature is held constant at $45 \
mK$. The dashed line shows the expectations from naive CF theory,
which predicts a $B^{-5/3}$ dependence until full polarization,
and an exponential increase in $T_1$ thereafter. As can be seen,
even without disorder, the predictions of self-consistent HF vary
substantially from the naive CF expectation. However, the
qualitative features of the naive expectation are indeed present
in the clean limit.
\begin{figure}[b]
\includegraphics[height=3in,width=3in]{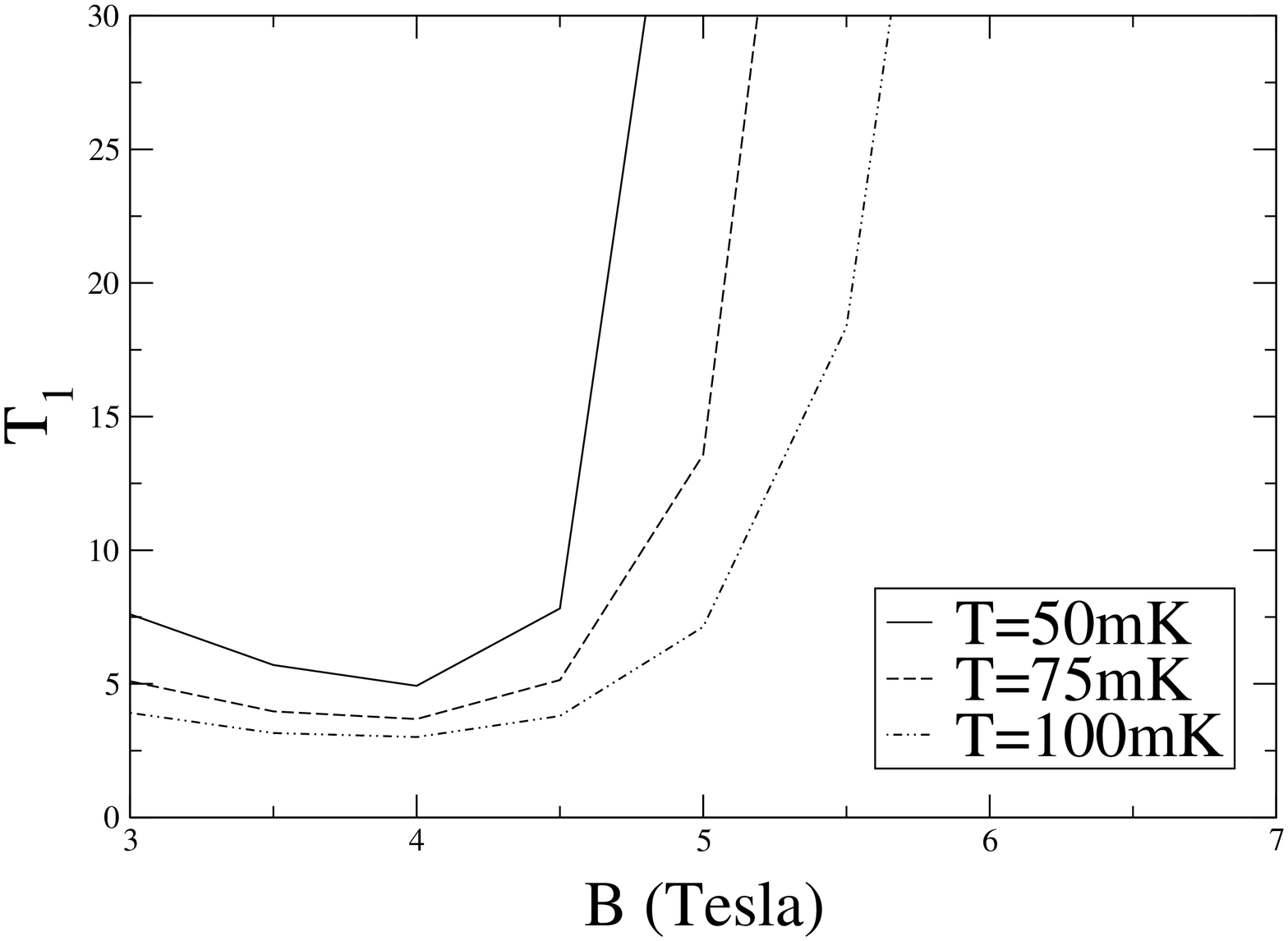}
\caption{The predictions of the HF approximation of the extended
Hamiltonian theory for the clean system for three different
temperatures. $T_1$ increases exponentially beyond full
polarization at the lowest temperature of $50mK$. Also note the
drop in $T_1$ below full polarization. } \label{fig5}
\end{figure}

The effect of disorder is to reduce the dependence on $B$ below
full polarization, and to make the rise after  full polarization
 gentler than exponential. The second feature can be
understood  by observing that even when the disorder-averaged
position of the bottom of the minority spin band is above the
chemical potential, the Lorentzian broadening of the band bottom
will produce some density of minority spin states at the chemical
potential. Assuming a constant $\Gamma$ this DOS can be seen to go
as $(B-B_c)^{-2}$, which implies a behavior $(B-B_c)^2$ for $T_1$.

In Fig. \ref{fig5} we show the behavior of $T_1$ in the clean system
($\G=0$) at three different temperatures. It is clear that at higher
temperatures, $T_1$ is lower due to a greater part of the band being
partially occupied. However, the rise of $T_1$ beyond full
polarization is extremely rapid, and it is clear that this is not
compatible with the experimental data\cite{TE}.

\subsection{Polarization}

\begin{figure}[t]
\includegraphics[height=3in,width=3in]{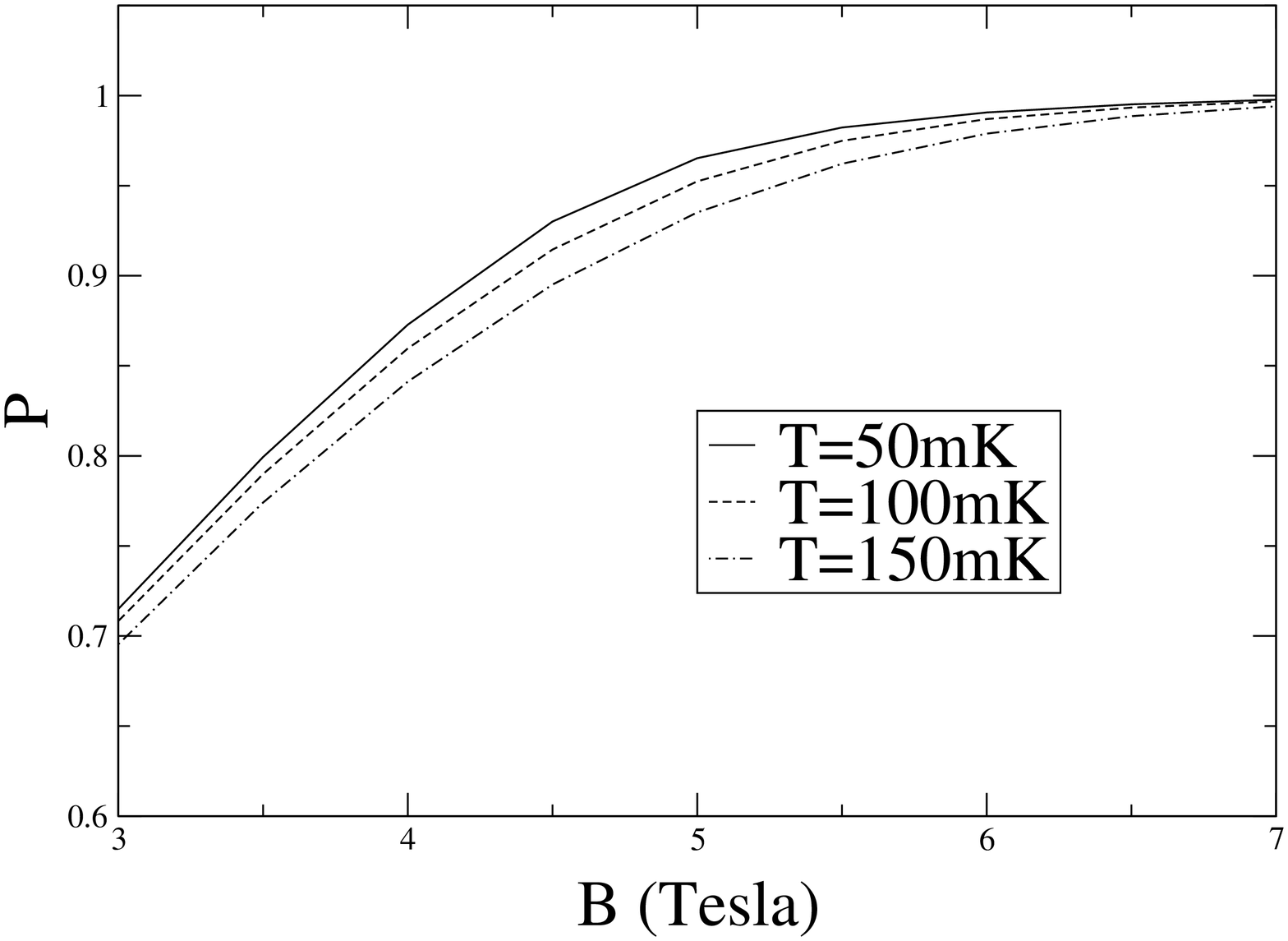}
\caption{The Polarization $P(B)$ for $\lambda=2.00$ and $\Gamma =
100mK$.} \label{fig6}
\end{figure}
As mentioned earlier, a compelling reason for preferring the
Hamiltonian theory over more phenomenological CF theories without a
direct link to the electronic Hamiltonian arises when one
simultaneously tries to explain data on independent quantities like
polarization and $1/T_1$. Where the simpler models call for mutually
incompatible values of mass and interaction parameters, the
Hamiltonian theory, given just $\lambda$ from one data point, is able
to account for all the salient features of the data.

To this end we are presenting here the predictions of our theory on
polarization $P(B,T)$ for the optimal values $\lambda = 2, \Gamma =
100mK$. Since the dependence of $P$ on $T$ is very weak for the range
considered by TE, we present just the $B$ dependence in Fig.
\ref{fig6}. Unfortunately TE do not measure the polarization. We
present these results in the hope that one day they may be tested.

\section{Qualitative effect of Landau-level mixing}

At fields of the order of a few Tesla the interaction scale
$e^2/\kappa l$ (approximately $50\sqrt{B} Kelvin$ where $B$ is in
$\ Tesla$) is comparable to the cyclotron scale $\hbar \omega_c$
(approximately $20 B Kelvin$ where $B$ is in $\ Tesla$). The ratio
${e^2\over\kappa l\hbar\omega_c}$ scales as $1/\sqrt{B}$. The
lower the field, the more important Landau-level mixing
is\cite{llmix-jim}. In this section we will show that the
qualitative effect of Landau-level (LL) mixing is to improve the
agreement between theory and experiment.

In previous work we have developed a formalism for taking Landau-level
mixing into account in our Extended Hamiltonian theory\cite{llmix-us},
where the electronic cyclotron coordinate $\etab_e$ is
retained. Briefly, by a unitary transformation we eliminate the
coupling to the higher Landau levels. To leading order, this results
in additional four-fermi and six-fermi\cite{llmix-us} terms
proportional to the perturbation parameter $\zeta={e^2\over\kappa
l\hbar\omega_c}$.

While this can be folded into the HF calculation we have described in
previous sections, it is computationally cumbersome. We will proceed
to take its effect into account qualitatively. LL-mixing produces
level repulsion between the LLL states and higher LL states. The
states at the top of the LLL are repelled more, with the net effect
being to reduce the interaction-induced bandwidth of the LLL. This
means that the density of states of CF's is increased by this effect,
implying that $T_1$ should decrease due to LL-mixing. This reduction
is larger at low fields and smaller at high fields. Applying this
insight to Fig. \ref{fig3} we see that LL-mixing will bring the
theoretical prediction into better agreement with the data.

The magnitude of this effect can be estimated by referring to our
previous calculations in the gapped fractions\cite{llmix-us}. For
$3 \ Tesla$, the LL-mixing parameter is roughly $\zeta=1.4$, while
for $6 \ Tesla$ is decreases to $1$. From Fig. 1 of Ref.
\cite{llmix-us} it is seen that the reduction of the FQH gap is
between 5-7\% in this range. We can thus expect the density of
states of the CF's to be enhanced a few percent at 3 $\ Tesla$
relative to the density of states at 6 $\ Tesla$. This will tend
to flatten the concavity of the theoretical curves at low $B$ and
improve the agreement with experiment.

\section{Conclusions, Caveats, and Open Questions}

In this paper we have focussed recent experiments of Tracy and
Eisenstein\cite{TE} on a $\nu =1/2$ system for temperatures as low as
$45mK$ and a range of magnetic fields.  They measured the nuclear
relaxation rate $1/T_1$ by magnetically disturbing the system and
measuring the relaxation of the resistance to its equilibrium
value. The predictions of the naive CF picture disagree strongly with
the measurements. We showed here that the extended Hamiltonian theory
developed by us gives a satisfactory account of the data if in
addition to the thickness parameter $\lambda$ of the electron-electron
interaction\cite{zds} we incorporate some constant, spin-independent,
disorder-induced width of order $\G=100mK$ for all momentum states.

Our Hartee-Fock calculation ignores vertex corrections due to gauge
fields\cite{read3,conserving-me} since they affect properties only
in the extreme low-energy/low-temperature limit. We also ignore
vertex corrections due to static disorder because the property we
calculate, $1/T_1$, is an integral over all momenta, and is
therefore expected to be insensitive to diffusion-like
contributions at small $\bq$. Our optimal value for the disorder
width $\Gamma$ is $100mK-120mK$, not too different from a crude
estimate based on an application of the Drude formula (Eq.
(\ref{drude})) to the measured longitudinal resistivity\cite{TE}.
This gives us confidence that our calculation is certainly an
important part of the explanation.

The value we have used for the thickness parameter in our
calculations, $\lambda=2.00$ at $B=4\ Tesla$, requires some
discussion. If one took the thickness parameter arising from the
self-consistent wavefunction in the transverse
direction\cite{zds,ando-fowler-stern} one would obtain $\lambda\simeq
1$. The primary effect of $\lambda$ is to soften the Coulomb
interaction at short distances. In {\it incompressible} states where
there is no linear screening, the finite thickness is indeed the
dominant contribution to softening the Coulomb interaction at short
distances. However, in a compressible system such as $\nu=\half$,
there will be further screening by the CF's themselves, which will
increase the effective value of $\lambda$ beyond that deduced from the
finite thickness of the 2DEG. We believe this is the reason our value
of $\lambda$ is so large.

Let us now turn to some caveats. The fundamental assumption underlying
our approach is that disorder averaging gives a true physical picture
of the system. While we expect this to be true when the system is far
from full polarization and the states of both spins are extended, we
expect that more complicated physics is relevant near full
polarization. In this regime, we can expect the minority spins to
become localized, perhaps in droplets at the minima of the disorder
potential. Because they are surrounded by a sea of majority spins, in
addition to the usual disorder potential, the minority spins feel an
exchange potential which tends to further localize them. Such effects
occur in zero-field Fermi liquids as well, but here the scale of the
exchange interactions is comparable to the Fermi energy, and exchange
effects are magnified. Thus it is doubtful whether disorder-averaging
will give a physically relevant answer. We can estimate the behavior
of $T_1$ for large $B$ in this regime as follows: Visualize droplets
of minority spin CF's filling the minima of the disorder potential. As
the Zeeman energy increases, the droplets will shrink such that their
area depends linearly on the Zeeman energy. Since the total relaxation
rate should be roughly proportional to the total area of such droplets
(relaxation requires that both spin species of CF's be present at a
given nuclear site in the localized regime) $1/T_1$ should decrease
linearly with the Zeeman energy. Actually, some droplets will
disappear when the Zeeman energy increases, so the decrease of $1/T_1$
will be faster than linear in $E_Z$, with the actual power depending
on the distribution of minima of the disorder potential. Note that the
disorder potential which should be used here is not the bare one due
to inhomogenieties of the dopants, but rather the bare potential
screened by the compressible CF state, and including exchange
contributions of the type mentioned earlier. A quantitative
calculation of these effects is beyond the scope of this paper. In
light of these arguments, it is rather fortuitous that our
disorder-averaged calculation seems to track the data
(Fig. \ref{fig2}) even beyond nominal full polarization.

Finally, we have made no attempt to construct a theory of the
longitudinal conductance\cite{hlr} in the LLL. As has been known for
some time\cite{currentLLmix}, the dominant terms in the current
operator are Landau-level mixing terms. However, it seems likely that
the low-frequency longitudinal conductance is controlled entirely by
LLL physics. An intriguing part of the data of TE\cite{TE} is the peak
in ${1\over\rho_{xx}}{d\rho_{xx}\over dE_Z}$ near nominal full
polarization. Our picture of the formation of localized droplets of
minority spin near nominal full polarization may have some relevance
for this peak as well. If such droplets are formed, they represent an
additional repulsive exchange potential for the {\it majority} spins,
which are mainly responsible for transport. Near nominal full
polarization, the system is extremely sensitive to the Zeeman energy,
and we expect the scattering due to the exchange potential of these
droplets to be maximized. We hope to revisit this issue in future
work.

\begin{acknowledgments}
We are grateful to Lisa Tracy and Jim Eisenstein for numerous
illuminating conversations and for generously sharing their data.
R. S. wishes to thank the NSF for partial support under the Grant
DMR-0354517.
\end{acknowledgments}


\begin{thebibliography}{1}
\bibitem{llmix-jim} We are indebted to Jim Eisenstein for encouraging
us in a private conversation to consider the effects of LL-mixing
on the relaxation rate.
\end{thebibliography}
\end{document}